\documentclass[ascmo, manuscript]{copernicus}

\begin{document}

\title{Analysis of the Evolution of Parametric Drivers of High-End Sea-Level Hazards}


\Author[1]{Alana}{Hough}
\Author[1]{Tony E.}{Wong}

\affil[1]{School of Mathematical Sciences, Rochester Institute of Technology, Rochester, NY 14623, USA}




\correspondence{Tony E. Wong (aewsma@rit.edu)}

\runningtitle{Parametric Drivers of Sea-Level Hazards}

\runningauthor{Hough and Wong}

\received{}
\pubdiscuss{} 
\revised{}
\accepted{}
\published{}


\firstpage{1}

\maketitle

    \begin{nolinenumbers}%

\begin{abstract}
Climate models are critical tools for developing strategies to manage the risks posed by sea-level rise to coastal communities.  While these models are necessary for understanding climate risks, there is a level of uncertainty inherent in each parameter in the models.  This model parametric uncertainty leads to uncertainty in future climate risks.  Consequently, there is a need to understand how those parameter uncertainties impact our assessment of future climate risks and the efficacy of strategies to manage them.  Here, we use random forests to examine the parametric drivers of future climate risk and how the relative importances of those drivers change over time.  We find that the equilibrium climate sensitivity and a factor that scales the effect of aerosols on radiative forcing are consistently the most important climate model parametric uncertainties throughout the 2020 to 2150 interval for both low and high radiative forcing scenarios.  The near-term hazards of high-end sea-level rise are driven primarily by thermal expansion, while the longer-term hazards are associated with mass loss from the Antarctic and Greenland ice sheets.  Our results highlight the practical importance of considering time-evolving parametric uncertainties when developing strategies to manage future climate risks.
\end{abstract}


\introduction  
A rising sea level poses a threat to island and coastal regions around the world.  More than 3.1 billion people globally live within 100 km of the coast \citep{coast_stats}.  Due to high populations of people in these regions, the respective governing bodies need to assess and manage risk \citep[e.g.,][]{climate_gov_doc_1, climate_gov_doc_2, climate_gov_doc_3, climate_gov_doc_4}.  Climate models provide a valuable tool to understanding future climate risks and testing the efficacy of risk management strategies in a computational experimental setting.

There are various modeling techniques that climate models are based on.   Semi-empirical models (SEMs) are both flexible and computationally efficient.  Because of that, they are appropriate for uncertainty quantification, resolving the high-risk upper tails of probability distributions, and informing decision analysis.  Other, more detailed process-based climate models (e.g., CLARA \citep{CLARA} or SLOSH \citep{SLOSH}) are also useful because they resolve more specific processes and have better geographical resolution. This work uses a SEM called the Building blocks for Relevant Ice and Climate Knowledge model \citep[BRICK;][]{BRICK}. 

However, climate models have numerous model parameters and multiple potentially conflicting data sets may be used to calibrate them, which poses a challenge when interpreting climate model outputs \citep{climate_model_uncertainties_1, climate_model_uncertainties_2}.  All models are an approximation of reality, so their parameters also have a level of parametric uncertainty. As a result, this parametric uncertainty contributes to uncertainty in the coastal hazard estimates presented to risk managers and decision-makers.  
  
With climate change comes changing risks, and there is a need for methods to both assess these risks and attribute their causes \citep{dynamic_adaptive_pathways_1, dynamic_adaptive_pathways_2, dynamic_adaptive_pathways_3}. To understand the impact on near-term and long-term risks, it is important to consider how the contribution of each parametric uncertainty to overall high-end sea-level hazard changes over time.  Understanding how these uncertainties change over time will aid in risk-averse decision-making related to adaptation to sea-level rise \citep{risk_aversion}.

The goal of this work is to understand how parametric uncertainties impact future climate risks.  We focus in particular on characterizing uncertainty in future high-end coastal hazards from mean sea-level rise.  Here, we accomplish this goal by using machine learning techniques.  Among the techniques we use are decision trees and random forests, which have previously been used in climate change studies.  For example, \citet{similar_work_1} used random forests to characterize the relative importance of uncertainties in relation to flooding due to sea-level rise.  \citet{similar_work_1} found that future coastal flood risk is driven by the uncertainty of human activities.  Similarly, \citet{similar_work_3} utilized random forests to assess regional flood risk.  When applied to the Dongjiang River Basin, China, \citet{similar_work_3} determined that maximum three-day precipitation, runoff depth, typhoon frequency, digital elevation model, and topographic wetness index are the most important risk indices.  \citet{similar_work_2} used random forests to analyze the impact of climate change on the wine regions of Hungary and found that in the long-term, only the northern region of Hungary will be suitable for the current grape crops that are typically grown in the country.

Decision trees are a supervised machine learning technique that successively splits a set of input data into different outcome regions.  Meanwhile, random forests are a collection of many decision trees constructed in a structured, but stochastic, way, designed to reduce overfitting to the training data.  To combat overfitting, random forests are created using a random subset of the training data and a random subset of the input features at each split. In this work, we use random forests to highlight which parameters impact the climate model projections the most and how each parameter’s impact changes over time.  \citet{Vega-Westhoff_paper} identified equilibrium climate sensitivity (ECS) as an important model parameter in a version of the BRICK model that is coupled to a carbon-cycle model.  In their work, they chose 5 $^{\circ}$C as the ECS cutoff that separates high-end and non-high-end sea-level risk scenarios.  Here, we can use random forests to examine what values of sea-level model parameters, including ECS, are most closely associated with high-end scenarios of sea-level rise.  We chose these machine learning methods due to their ability to process large amounts of climate model output and climate data to determine each parameter’s impact on future climate hazards.

\section{Methods}

\subsection{Models and data}
We use the model output from the coupled Hector-BRICK model \citep{Vega-Westhoff_paper}.  Hector is a simple climate carbon-cycle model \citep{hector}.  Similarly, the Building blocks for Relevant Ice and Climate Knowledge (BRICK) model is a simple climate model for simulating global mean surface temperature and global mean sea-level rise, as well as regional sea-level rise \citep{BRICK}.  BRICK uses the Diffusion-Ocean-Energy balance CLIMate model \citep[DOECLIM;][]{DOECLIM}, the Glaciers and Ice Caps portion of the MAGICC climate model \citep[GIC-MAGICC;][]{GIC-MAGICC}, a Thermal Expansion (TE) module based on the semi-empirical relationships used by (e.g.) \citet{TE_1} and \citet{TE_2}, the Simple Ice-sheet Model for Projecting Large Ensembles \citep[SIMPLE;][]{SIMPLE}, the ANTarctic Ocean temperature model \citep[ANTO;][]{ANTO}, and the Danish Center for Earth System Science Antarctic Ice Sheet model \citep[DAIS;][]{DAIS}.  Since our main focus is on changes in global mean sea level (GMSL), it is important to note that BRICK determines this by summing the sea-level change due to changes in land water storage \citep{land_water_storage}, glaciers and ice caps, the Greenland ice sheet, the Antarctic ice sheet, and thermal expansion.  

We use the model output data from the Representative Concentration Pathway \citep[RCP;][]{RCP} 2.6 and 8.5 scenarios from BRICK simulations of \citet{Vega-Westhoff_paper}.  Comprehensive discussions of the Hector model is given in \citet{hector}, of the BRICK model in \citet{BRICK}, and of the Hector-BRICK coupled model in \citet{Vega-Westhoff_paper}. This work focuses on the BRICK model for sea-level rise and how uncertainty in its parameters relate to uncertainty in future coastal hazards, so we use only the BRICK model parameters and their relation to the sea-level rise scenarios. A list of the 38 parameters of the BRICK model are shown in Table \ref{param_table}. These model parameters include, but are not limited to, equilibrium climate sensitivity (ECS), a factor that scales the effect of aerosols on radiative forcing ($\alpha_{DOECLIM}$), thermal expansion ($\alpha_{TE}$), the temperature associated with the onset of fast dynamical disintegration of the Antarctic ice sheet ($T_{crit}$), and the rate of fast dynamical disintegration of the Antarctic ice sheet ($\lambda$) \citep{AIS_disintegration_3}.

We use the model output from \citet{Vega-Westhoff_paper} that has been calibrated using observations of global mean surface temperature and sea-level rise due to thermal expansion, glaciers and ice caps, the Greenland ice sheet, and the Antarctic ice sheet to constrain model parameters and projections of future sea levels and temperatures. We use the 10,000 parameter sample values from the “TTEGICGISAIS.csv” file along with projected GMSL values for the 2020 to 2150 time period \citep{vega_westhoff_data}. This corresponds to the results presented in the main text of that work. The original data from \citet{vega_westhoff_data} were converted from their original RData file format to CSV using R version 3.6.1 \citep{R}.  The subsequent analyses and plots were done in Python \citep{python}, using the sklearn library to make the decision trees and random forests \citep{scikit-learn}.

The aim of the present work is to explore high-end sea-level rise scenarios and analyze how the factors driving sea-level hazards change over time. Toward this end, we preprocessed the data.  The GMSL model output from every five years between 2020 and 2150 was our output data of interest.  We went through each year of GMSL outputs in those 5-year intervals from the different RCP scenarios and calculated the 90th percentile of the GMSL ensemble.  We used the 90th percentile as the threshold for classifying “high-end” scenarios of sea-level rise as any state-of-the-world (SOW, or ensemble member) that meets or exceeds this value; a concomitant set of model parameters, RCP forcing, and resultant temperature change and GMSL change comprises a SOW.  SOWs with GMSL in each target year below this threshold are classified as “non-high-end.” It is possible for a SOW to have non-high-end GMSL in one 5-year time period and later have high-end GMSL.  In addition to the 90th percentile threshold, we considered the 80th percentile as the threshold in a supplemental experiment, and found that the results were not sensitive to the selection of the percentile threshold (see Appendix).

\subsection{Decision trees}
We are interested in examining how a given SOW’s model parameters are related to whether that SOW is more/less likely to be a high-end scenario of GMSL. Decision trees are a supervised learning approach used in classification and regression applications \citep{ISL_textbook}.  We use decision trees to classify each set of model parameters as leading to high-end or non-high-end sea-level rise by successively splitting training outcomes into different outcome regions.

As a running example, Fig. \ref{decision_tree_ex} shows a graphical representation of a hypothetical decision tree.  Figure \ref{decision_tree_ex} splits on equilibrium climate sensitivity (ECS), and the parameters $P_0$ and $b_{SIMPLE}$.  ECS is defined as the equilibrium increase in global mean surface temperature that results from doubling the atmospheric CO$_2$ concentration relative to pre-industrial conditions, and is related to the climate component of the BRICK model.  Meanwhile $P_0$ and $b_{SIMPLE}$ pertain to the major ice sheets’ components of the model.  $P_0$ represents the Antarctic annual precipitation for Antarctic surface temperature of 0 $^{\circ}$C, and $b_{SIMPLE}$ is the equilibrium Greenland ice sheet volume for temperature anomaly of 0 $^{\circ}$C.

Once the levels of splits in a tree reach a specified depth or another specified stopping criteria, the outcome from each leaf node is determined.  The following are decision tree hyperparameters that can be used as stopping criteria in the sklearn library:
\begin{itemize}
    \item max\_depth, the maximum allowed depth of a tree;
    \item min\_samples\_leaf, the minimum number of samples needed at a node in order for it to be a leaf node; and
    \item min\_samples\_split, the minimum number of samples needed at a node in order for a split to occur \citep{scikit-learn}.
\end{itemize}
If the number of samples at a node is less than min\_samples\_split, then the node will be a leaf node.

In the example depicted in Fig. \ref{decision_tree_ex}, the maximum depth (max\_depth) we use is 3, so the tree splits on three levels before creating leaf nodes.  In our case, the leaf nodes would be classified as “non-high-end” or “high-end” as described above by a simple majority vote among the data points allocated to that node.  The values for the parent split nodes are determined by considering the information gain of each possible split.  Information gain quantifies the reduction in impurity, such as entropy, that would occur as a result of that split.  A large value of information gain is desirable.  Therefore, the potential parameter choice and value for that parameter that give the largest information gain will be selected as the split. 

After this training procedure, in which data from the ensembles of \citet{Vega-Westhoff_paper} is used to determine the split node values and leaf outcome classifications, the decision tree can predict outcomes based on input feature data.  For example, a feature data point $x$ with ECS equals 5 $^{\circ}$C, $P_0$ equals 0.2 m yr$^{-1}$, and $b_{SIMPLE}$ equals 9 m would be classified as “high-end.”  Starting at the top of the tree, we consider the ECS value.  Since $x$ has an ECS greater than 3.25 $^{\circ}$C, we move down the right branch of the tree to the $P_0$ node.  $x$’s $P_0$ value, 0.2 m yr$^{-1}$, is less than 0.5 m yr$^{-1}$, so we continue to the left child of the $P_0$ node, which is the $b_{SIMPLE}$ node.  Because $x$ has $b_{SIMPLE}$ value greater than 7.9 m, we go to the right child of the $b_{SIMPLE}$ node.  This node is a “high-end” leaf node, so we classify $x$ as “high-end.”
\begin{figure*}[t]
    \includegraphics[width=12cm]{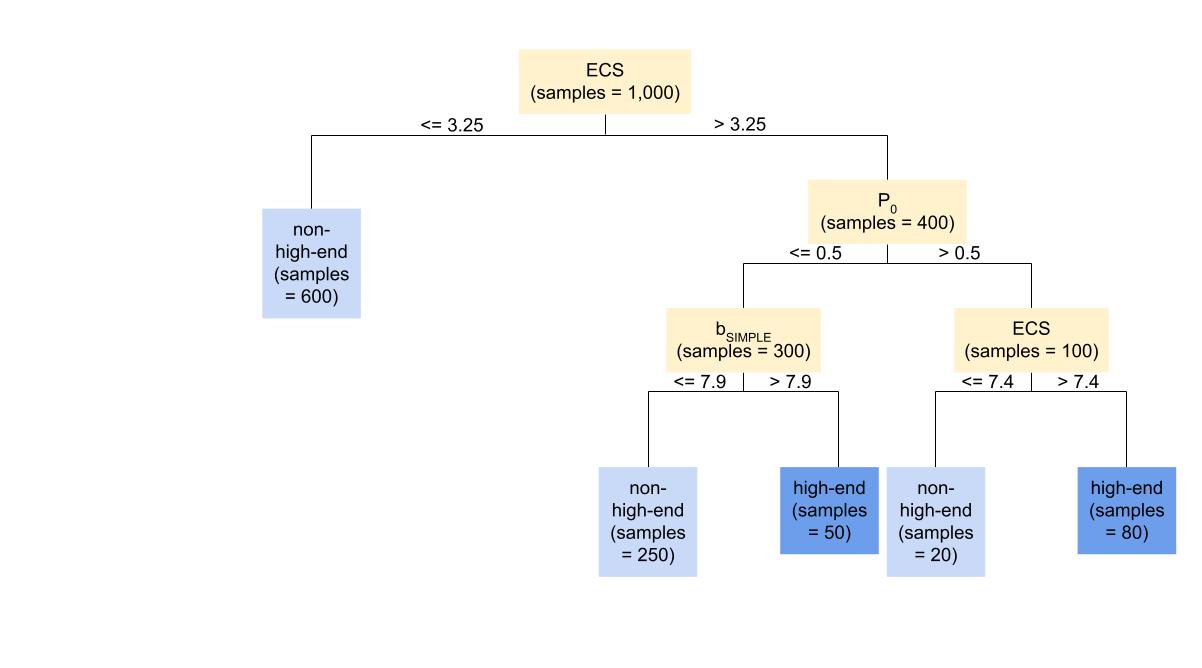}
    \caption{Hypothetical decision tree demonstrating the general decision tree structure using BRICK model parameters and our “high-end” and “non-high-end” classification outcomes.  Since we used a maximum depth of 3 as the stopping criteria, the tree made three levels of splits before stopping to create leaf nodes.}
    \label{decision_tree_ex}
\end{figure*}

\subsection{Random Forests}
As can be the case with many machine learning algorithms, decision trees can overfit the training data used to create the tree \citep{ISL_textbook}.  Random forests, which are an ensemble method, are one way to reduce overfitting. Random forests are a collection of many decision trees, created by using a random subset of the training data to build each tree and by using a random subset of the features at each split.  In this work, the features of the random forests are the parameters of the BRICK model, and the response to be classified is whether or not the time series of GMSL associated with those parameters is a high-end GMSL scenario (above the 90th percentile).  Taking a random subset of the training data when creating a tree is called bootstrapped aggregation, or bagging.  When bagging, the random subset of the training data is taken with replacement.  In addition to bootstrap subsets of training data, random forests take a random subset of features from which to split the data.  This further helps to address the issue of overfitting the training data.

The following general process outlines how we construct each random forest.  We repeat this process to classify the high-end GMSL scenarios in 5-year increments from 2020 through 2150 for each of RCP2.6 and RCP8.5. This leads to a total of 27 random forests for each of the two RCP scenarios.

To create a random forest of decision trees, we first split the data into the parameters and the output for the given year.  We then created training and test subsets of that data.  The training set is used to train the model and tune the model’s hyperparameters, and the test set is used to test the model’s performance.  In this work, the training subset comprises 80\% of the original data, and the test comprises 20\%.  We use the RandomForestClassifier from the sklearn.ensemble library to make a forest of decision trees using entropy as the criterion to determine the splits of each tree \citep{scikit-learn}.

Prior to fitting the model, we tune the hyperparameters of the RandomForestClassifier by performing a five-fold cross validation grid search.  This process takes the training dataset, divides it into five equal-sized subsets (or “folds”), and uses four subsets to train the model and one subset as the validation set.  This is repeated five times, so that each fold is used once as the validation set. The same parameters are used to train random forests with each combination of four training subsets and one validation subset.  Therefore, the same parameters are used with five different training sets, and this is done for each combination of hyperparameters in the grid.   In addition to the three decision tree-specific hyperparameters noted in Sect. 2.2, we explored values for the following parameters in our grid search: 
\begin{itemize}
    \item max\_features, the number of features a tree can consider at each split; and 
    \item n\_estimators, the numbers of trees in the forest \citep{scikit-learn}. 
\end{itemize}

We tried a few preliminary grid searches over these hyperparameters to narrow down the ranges for the hyperparameters before the final grid search using the ranges shown in Table \ref{gridsearch_values_table}.  To select hyperparameters for all of the main experiments, we used these ranges in grid searches for the years 2050, 2075, and 2100 for both RCP2.6 and RCP8.5.  The cross-validation scores for the hyperparameter sets for a given RCP and year are very similar (Table \ref{2050_hyperparameter_table}, Table \ref{2075_hyperparameter_table}, Table \ref{2100_hyperparameter_table}), and therefore, are not sensitive to changing the hyperparameters for the given ranges in Table \ref{gridsearch_values_table}.  We chose the set of best values in Table \ref{gridsearch_values_table} to use as the hyperparameter values because that specific combination of values consistently had one of the top cross-validation scores in the years and RCP scenarios that we did the gridsearch over (Table  \ref{2050_hyperparameter_table}, Table \ref{2075_hyperparameter_table}, Table \ref{2100_hyperparameter_table}).

\begin{table*}[t]
    \caption{Hyperparameter values of the best estimator from the gridsearch.  The max\_features hyperparameter is fixed at ‘sqrt’ to ensure variability across the decision trees within the random forest.  The ‘sqrt’ value for max\_features means the square root of the total number of features will be used as the numerical value for max\_features.}
    \begin{tabular}{c|c|c|c|c|c}
        \textbf{Hyperparameter} & max\_depth & max\_features & min\_samples\_leaf & min\_samples\_split & n\_estimators \\ \hline
        \textbf{Values searched over} & 12, 14, 16, 18 & 'sqrt' & 4, 7, 10, 13 & 7, 10, 13, 16 & 250, 500, 750, 1000 \\ 
        \textbf{Best value} & 16 & 'sqrt' & 4 & 7 & 500 \\ 
    \end{tabular}
    \label{gridsearch_values_table}
\end{table*}

Using the random forest hyperparameters settings from the “Best value” row of Table \ref{gridsearch_values_table}, we fit random forests using the training data for each of the 5-year intervals for each of RCP2.6 and 8.5.  We input the corresponding testing parameter values into the random forest models for it to predict output values.  We then compare the predicted values to the actual output values from the testing data and calculate the percentage of the testing values that the model correctly predicted.  The same process is done for the training subset that was used to create the forest.  The training and testing accuracies for the forests can be found in Table \ref{forest_accuracy}.

Given the initial hypothesis that the equilibrium climate sensitivity (ECS) parameter is an important indicator of future sea-level rise \citep{Vega-Westhoff_paper}, we examine the set of all split values for the ECS parameter for each tree in the forest. Likewise, we examine the distribution of the maximum ECS value of each tree in the forest.  The maximum split values are informative because they differentiate the highest cases of sea-level rise when the trees are branching.  Hence, the maximum splits help quantify the threshold of ECS that separates the high-risk situations from the non-high-risk (lower) ECS splits.

\subsection{Feature Importances}
We use feature importances to assess which parameters play the largest role in determining whether or not a given SOW is a high-end GMSL scenario.  We compute the importances for each feature (i.e., BRICK model parameter) using our forests fit for each year from 2020 to 2150 in 5-year increments.  We use Gini importances as the feature importances, which are the normalized reductions of node impurity by each feature chosen for splitting the tree \cite{scikit-learn}.  We calculate the importance of each node in a tree using Eq. (\ref{eq_1}).
\begin{equation}
    \label{eq_1}
    ni_j = w_jC_j - w_{\text{left}, j}C_{\text{left}, j} - w_{\text{right}, j}C_{\text{right}, j}
\end{equation}

$ni_j$ is the node importance of node $j$, $w_j$ is the weighted number of samples at node $j$, $C_j$ is the impurity of node $j$, the “left” subscript represents the left child from the split on node $j$, and the “right” subscript represents the right child from the split on node $j$.  The weighted number of samples is used as a coefficient of the different impurity calculations because the impurities of nodes address the proportion of data points that belong to that node’s left- and right-children, but does not address the total number of data points associated with that node. The calculation of $ni_j$ in Eq. (\ref{eq_1}) addresses this by giving greater weight to nodes with a large proportion of the samples than nodes that split small numbers of samples.

For example, in Fig. \ref{decision_tree_ex}, the root node is the ECS split with a value of 3.25 $^{\circ}$C.  Considering that node, the node importance would be Eq. (\ref{eq_2}).
\begin{equation}
    \label{eq_2}
    ni_{\text{ECS}, 3.25} = 1000C_{\text{ECS}, 3.25} - 600C_{\text{left}; {\text{ECS}, 3.25}} - 400C_{\text{right}; {\text{ECS}, 3.25}}
\end{equation}

The impurity of a node is a measure of the efficacy of the feature used for splitting the data set at that node for subdividing the data set.  We use entropy as the impurity criterion in our forests, which ranges in value from 0 to 1.  If the data at a given node are split evenly among that node’s left- and right-children, then the node impurity is maximized at 1. The more asymmetrically the data at that node are split between the node’s children, the lower the node’s entropy will be. The closer the entropy is to 1, the more difficult it is to draw conclusions from the data split by that node.  We calculate entropy using Eq. (\ref{eq_3}, where $p_c$ is the fraction of examples in class $c$ (where $c$ here is either high-end GMSL or non-high-end GMSL).
\begin{equation}
    \label{eq_3}
   C_j = entropy = - \sum_c p_c log_2(p_c)
\end{equation}

In the example using Fig. \ref{decision_tree_ex}, the entropies would be the following:
\begin{align}
    C_{\text{ECS}, 3.25} & = - (p_{\text{high-end}} log_2(p_{\text{high-end}}) + p_{\text{non-high-end}} log_2(p_{\text{non-high-end}})) \\
    & = - \left(\frac{130}{1000} log_2\left(\frac{130}{1000}\right) + \frac{870}{1000} log_2\left(\frac{870}{1000}\right)\right) \\
    & \approx 0.5574
\end{align}
\begin{align}
    C_{\text{left}; {\text{ECS}, 3.25}} & = - (p_{\text{high-end}} log_2(p_{\text{high-end}}) + p_{\text{non-high-end}} log_2(p_{\text{non-high-end}})) \\
    & = - \left(\frac{0}{600} log_2\left(\frac{0}{600}\right) + \frac{600}{600} log_2\left(\frac{600}{600}\right)\right) \\
    & = 0
\end{align}
\begin{align}
    C_{\text{right}; {\text{ECS}, 3.25}} & = - (p_{\text{high-end}} log_2(p_{\text{high-end}}) + p_{\text{non-high-end}} log_2(p_{\text{non-high-end}})) \\
    & = - \left(\frac{130}{400} log_2\left(\frac{130}{400}\right) + \frac{270}{400} log_2\left(\frac{270}{400}\right)\right) \\
    & \approx 0.9097
\end{align}

With these, the node impurity of the ECS split with a value of 3.25 $^{\circ}$C can be fully calculated as shown in Eq. (\ref{eq_13}), Eq. (\ref{eq_14}), and Eq. (\ref{eq_15}).
\begin{align}
    \label{eq_13}
   ni_{\text{ECS}, 3.25} & = 1000C_{\text{ECS}, 3.25} - 600C_{\text{left}; {\text{ECS}, 3.25}} - 400C_{\text{right}; {\text{ECS}, 3.25}} \\
   \label{eq_14}
   & = 1000(0.5574) - 600(0) - 400(0.9097) \\
   \label{eq_15}
   & = 193.52 
\end{align}

Once the node importances of a tree are calculated, we use them to calculate feature importances in Eq. (\ref{eq_16}).
\begin{equation}
    \label{eq_16}
    fi_i = \dfrac{\sum_{j: \: node \: j \:splits \: on \: feature \: i} ni_j}{\sum_{n \in all \: nodes} ni_n}
\end{equation}

In Eq. (\ref{eq_16}), $fi_i$ is the feature importance of feature $i$.  The feature importances are normalized so they sum to 1.  Equation (\ref{eq_17}) demonstrates normalizing the previously calculated feature importances.
\begin{equation}
\label{eq_17}
    fi_{i, normalized} = \dfrac{fi_i}{\sum_{k \in all \: features} fi_k}
\end{equation}

Since we are constructing forests of trees, the feature importances that we present for a given forest are the mean feature importances over all the trees in the forest.

We compute the feature importances for each feature for each forest that we fit, in 5-year intervals from 2020 to 2150.  With these importances, we construct a stacked bar graph from 2020 to 2150 in 5 year increments.  The bar for each year shows the breakdown of the feature importances for that specific year.  Hence, all of the individual feature importances bars for a given year will add up to 1.  We define an “other” category such that model parameters with an importance less than 4\% are grouped into “other”.  There are 38 model parameters, so if the importances were uniform across all the parameters, each importance would be about 2.6\%.  Hence, any “other” category threshold of 3\% or less would show importances that were not substantially different from the average.  Because of that, we use 4\% as the threshold for the “other” category.

\section{Results}
\subsection{Feature Importances}
Based on Fig. \ref{importances_plot}, equilibrium climate sensitivity (ECS) (darkest solid blue boxes) and the aerosol scaling factor ($\alpha_{DOECLIM}$) (darkest stippled blue boxes) are consistently associated with greatest high-end sea-level risk throughout both RCP2.6 and RCP8.5.  Both of these model parameters are associated with the climate component of the BRICK model.  Equilibrium climate sensitivity (ECS) and the aerosol scaling factor ($\alpha_{DOECLIM}$) account for 9.3\% and 6.4\% (respectively) of the overall feature importance in RCP2.6 in the year 2020 and increase to 28.1\% and 15.9\% by the year 2150 (Fig. \ref{importances_plot}a). In the higher forcing RCP8.5 scenario, ECS accounts for 9.2\% and $\alpha_{DOECLIM}$ accounts for 6.6\% of the overall feature importance in 2020.  By 2150 in RCP8.5, they increase to 13.6\% and 12.0\% respectively (Fig. \ref{importances_plot}b). The relatively higher importance associated with these climate module parameters in the lower forcing RCP2.6 scenario are indicative of the large influence exerted by those parameters on the severity of the resulting sea-level rise. The only parameter in RCP2.6 that has an importance greater than 4\% and belongs to a non-climate component of BRICK is the temperature associated with Antarctic ice sheet fast disintegration, $T_{crit}$ (Fig. \ref{importances_plot}a). This suggests that in the high-end SOW of sea-level rise, even in the low-forcing RCP2.6 scenario, by the middle of the 21st century and beyond the Antarctic ice sheet dynamics can still drive severe risks to coastal areas.
\begin{figure*}[t]
    \includegraphics[width=12cm]{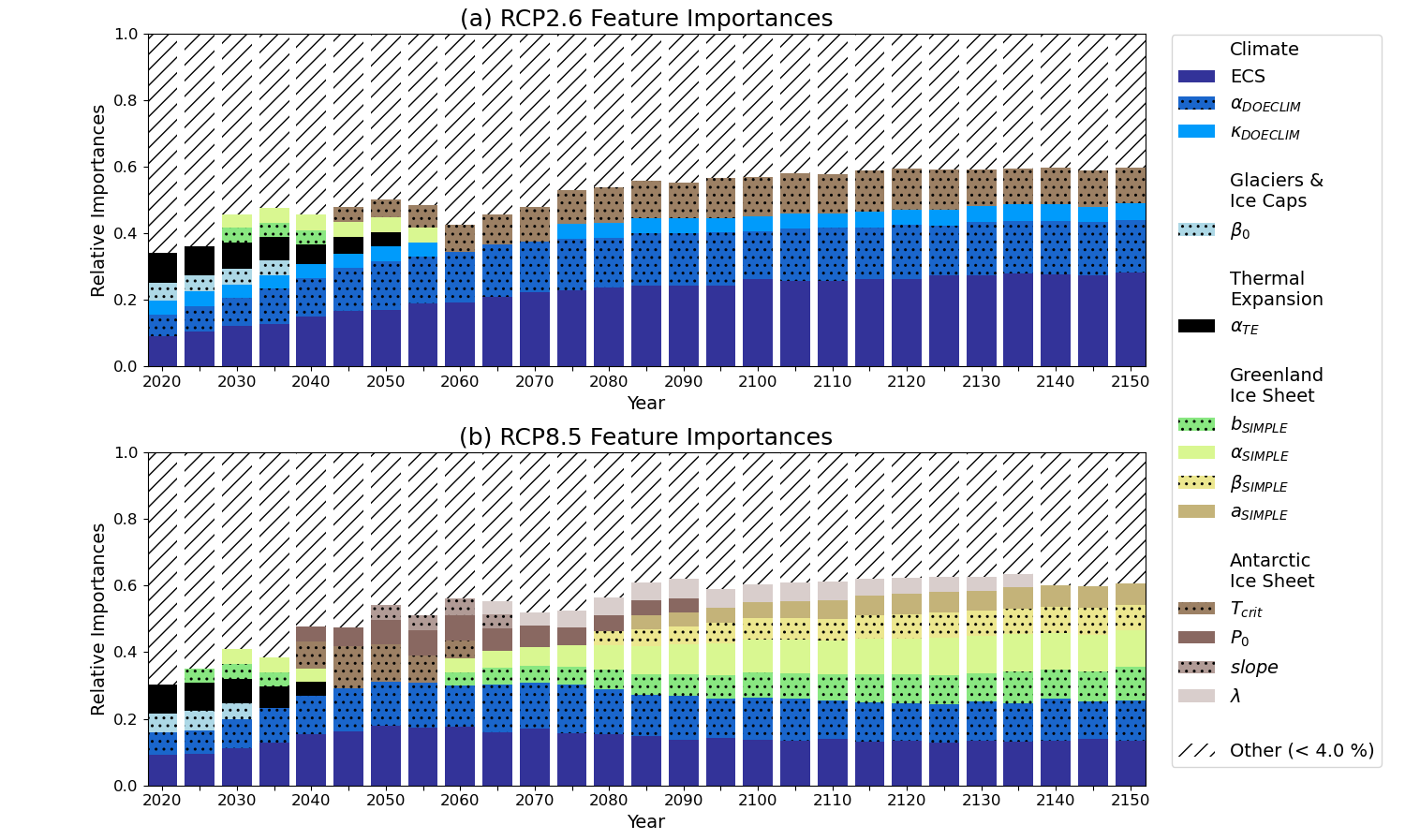}
    \caption{Relative feature importances of the BRICK model parameters calculated based on the fitted random forests. Shown are the importances of the BRICK model parameters using (a) the RCP2.6 radiative forcing scenario, and (b) the RCP8.5 forcing scenario. All model parameters with an importance less than 4\% were grouped into an “other” category, which is shown with hatch marks to denote its difference from the parameters.  Stippling was added to alternating parameters in the legend to aid in telling the difference between similar colors.}
    \label{importances_plot}
\end{figure*}

\subsection{Characterization of risk over time}
In contrast to high-end sea-level rise being driven primarily by climate uncertainties under RCP2.6, in the higher forcing RCP8.5 scenario, the importances of the ECS and aerosol scaling factor are relatively lower. This is indicative of the transition to uncertainty in sea-level processes driving high-end risks under the higher forcing scenario. The near-term risk in both RCP2.6 and RCP8.5 is driven by sea-level rise from thermal expansion ($\alpha_{TE}$).  Figure \ref{importances_plot} (solid black boxes) shows that the parameters related to thermal expansion are important from 2020 until the middle of the century, becoming less important as time goes on.  In RCP2.6, $\alpha_{TE}$ comprises 9.0\% of the overall feature importance in 2020, which then decreases to 4.4\% in 2050.  Likewise in RCP8.5, $\alpha_{TE}$ accounts for 8.8\% in 2020 and decreases to 4.3\% in 2040.

As for the long-term risks, ice loss from the Antarctic ice sheet is a driver in both emissions scenarios.  In particular, the $T_{crit}$ and $\lambda$ parameters within the Antarctic ice sheet component of BRICK are important (Fig. \ref{importances_plot}, stippled dark brown boxes and solid taupe respectively).  Within the BRICK model, $T_{crit}$ is the temperature associated with the onset of fast dynamical disintegration of the Antarctic ice sheet.  In RCP8.5, $T_{crit}$ is significantly important from 2040 to 2060.  This is consistent with predictions that major Antarctic ice sheet disintegration will occur between 2040 and 2070 in high radiative forcing scenarios \citep{AIS_disintegration_1, AIS_disintegration_2, AIS_disintegration_3, AIS_disintegration_4}.

In addition to the Antarctic ice sheet, ice loss from the Greenland ice sheet also poses a risk in the long term.  In the case of the Greenland ice sheet, the sea-level rise associated with its ice loss only has significant importance in RCP8.5.  The specific model parameters associated with the Greenland ice sheet are the following: $b_{SIMPLE}$, $\alpha_{SIMPLE}$, $\beta_{SIMPLE}$, and $a_{SIMPLE}$ (Fig. \ref{importances_plot}b; stippled green boxes, solid light green boxes, stippled yellow boxes, and solid tan boxes respectively).

\subsection{ECS threshold of high-end sea-level rise}
Previous work by \citet{Vega-Westhoff_paper} used 5 $^{\circ}$C as the value of ECS that separates high-end and non-high-end climate risk scenarios.  Using our collection of random forests, we select the highest ECS split value that each decision tree in the forest split on. These split values should separate the highest-risk scenarios of GMSL in each time interval. Figure \ref{ECS_boxplot} shows the distribution of the maximum ECS splits for every 5 years in the 2020 to 2150 time period.
\begin{figure*}[t]
    \includegraphics[width=12cm]{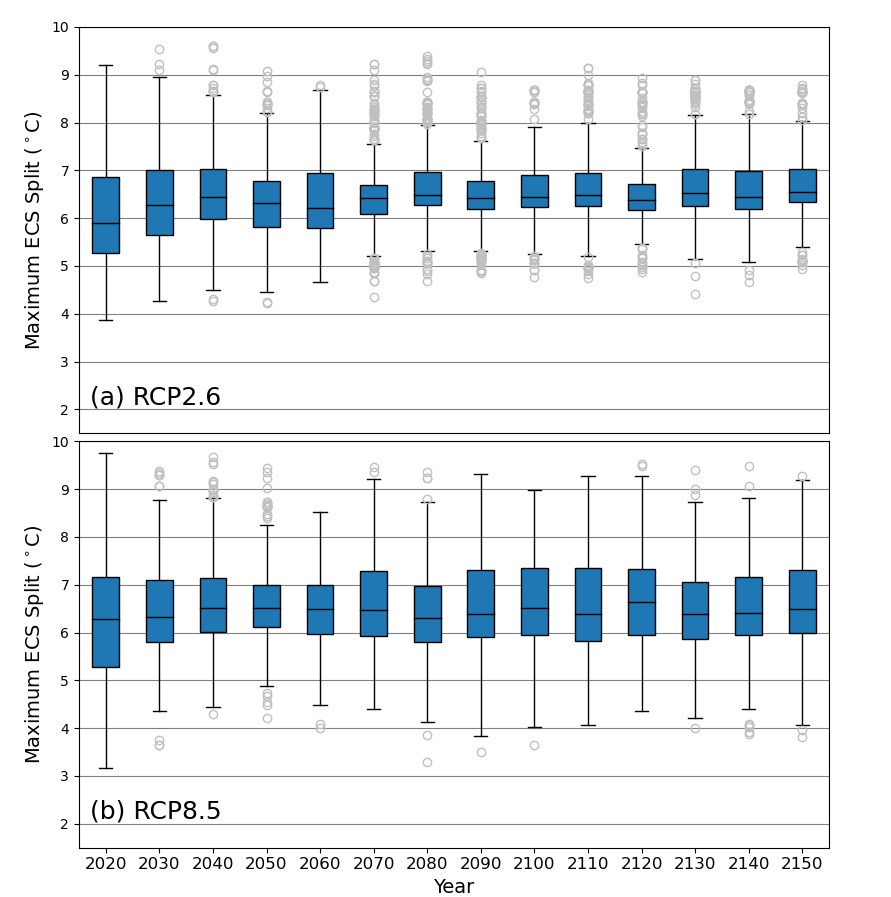}
    \caption{Distributions of the maximum equilibrium climate sensitivity (ECS) split value from each decision tree in the fitted random forests.  (a) depicts the maximum ECS split distributions in the RCP2.6 forests, and (b) depicts the maximum ECS split distributions in the RCP8.5 forests.  The outliers are the data points less than $\text{Q1} - 1.5* \text{IQR}$ or greater than $\text{Q3} + 1.5* \text{IQR}$, where Q1 is Quartile 1, Q3 is Quartile 3, and IQR is the interquartile range (Q3$-$Q1).  The blue boxes show the IQR, and the line within the IQR is the median.}
    \label{ECS_boxplot}
\end{figure*}

There is a considerable number of outliers of the maximum ECS split value in RCP2.6, particularly in the high-risk upper tail (Fig. \ref{ECS_boxplot}a). However, the interquartile ranges are quite small. In most cases, the interquartile range for RCP2.6 falls between approximately 5.75 $^{\circ}$C and 7 $^{\circ}$C (Fig. \ref{ECS_boxplot}a).  The greatest interquartile range width is 1.6 $^{\circ}$C in 2020, and the smallest width is 0.5 $^{\circ}$C in 2120, with an average of 0.9 $^{\circ}$C (Table \ref{max_ECS_split_quartiles}).   Likewise, the median is consistently between 6 $^{\circ}$C and 6.5 $^{\circ}$C.  In RCP2.6, the medians of the maximum ECS split over the given years ranges from 5.9 $^{\circ}$C to 6.6 $^{\circ}$C (Table \ref{max_ECS_split_quartiles}).

There are fewer outliers in RCP8.5 than in RCP2.6, but RCP8.5 has wider uncertain ranges.  The interquartile ranges of the maximum ECS split value in RCP8.5 span a larger length than those of RCP2.6.  Here, the interquartile ranges are mostly contained between 5.75 $^{\circ}$C and 7.25 $^{\circ}$C  (Fig. \ref{ECS_boxplot}b).  The interquartile range width in RCP8.5 is between 0.9 $^{\circ}$C (in 2050) and 1.9 $^{\circ}$C (in 2020), with an average of 1.3 $^{\circ}$C (Table \ref{max_ECS_split_quartiles}).  However, the median is consistently between 6 $^{\circ}$C and 6.5 $^{\circ}$C.  In RCP8.5, the medians of the maximum ECS split over the given years ranges from 6.29 $^{\circ}$C to 6.64 $^{\circ}$C (Table \ref{max_ECS_split_quartiles}).

Since the median is consistently between 6 $^{\circ}$C and 6.5 $^{\circ}$C in both low and high radiative forcing scenarios, our random forests suggest that an ECS value between 6 $^{\circ}$C and 6.5 $^{\circ}$C could be used as a threshold for high-end scenarios of global mean sea-level rise. In a supplemental experiment, we reproduced the same experiment using the 80th percentile of the global mean sea level to denote the high-end cases (as opposed to the 90th percentile in the main experiments).  We find that the distribution of the maximum ECS split value from each decision tree in the random forests fitted using the 80th percentile of the data is very similar to that of the 90th percentile (Fig. \ref{80th_percentile_max_ECS_boxplot}). Hence, our results are not sensitive to our choice to use the 90th percentile.

\conclusions  
In this work, we present an approach to classify global mean sea level from the BRICK semi-empirical model as either high-end or non-high-end.  We use a previously published model output  data set to construct random forests for 5 year increments from 2020 to 2150 for RCP2.6 and RCP8.5.  We explore the parametric drivers of risk by examining the feature importances of random forests.  Results show that climate components of the BRICK model, specifically equilibrium climate sensitivity (ECS) and the aerosol scaling factor ($\alpha_{DOECLIM}$), are consistently the most important parametric uncertainties in both radiative forcing scenarios (Fig. \ref{importances_plot}).  Our results also show that thermal expansion and glacier and ice cap mass loss both pose a risk in the near-term, and that long-term risks are driven by mass loss from the Greenland and Antarctic ice sheets (Fig. \ref{importances_plot}).  In addition to the feature importances, we find that an ECS value greater than 6 $^{\circ}$C to 6.5 $^{\circ}$C indicates a high-end sea-level rise scenario based on  the maximum ECS split values that each decision tree in a forest split on (Fig. \ref{ECS_boxplot}).  This result is consistent for both RCP2.6 and RCP8.5, as well as when using the 80th or 90th percentile to characterize the sea-level data as ``high-end''.

These results demonstrate the nonstationary risks posed by climate change and the related hazards driven by sea-level change. In turn, climate risk management strategies must address both near-term actions to mitigate near-term risks such as sea-level rise from thermal expansion and glaciers and ice caps.  At the same time, risk management strategies must also guard against the long-term risks driven by mass loss from the major ice sheets.  While this work was centered around the impact of model parametric uncertainties on sea-level hazards, the same machine learning approaches can be generalized to incorporate the socioeconomic uncertainties that relate future climate hazards (e.g., changes in temperatures and sea levels) to financial and human risks.  These approaches offer promise to provide a more holistic view of uncertainties affecting future climate risk and the efficacy of human strategies to mitigate and manage these risks.




\codedataavailability{All codes are freely available from \url{https://drive.google.com/drive/folders/1iHSPr7qpBcQDwrKLQyem8o10duvaMNAG?usp=sharing}.  All data and modeling and analysis codes are freely available from \url{https://drive.google.com/drive/folders/1iHSPr7qpBcQDwrKLQyem8o10duvaMNAG?usp=sharing}. \textit{(After the review process, these Google Drive links will be replaced by a stable Zenodo repository URL and this italicized statement will be deleted.)}
} 



\appendix
\section{}    
\begin{table}[t]
    \caption{Hector-BRICK model parameter names, units, and descriptions \citep{supplemental_parameter_table}}
	\label{param_table}
	\begin{tabular}{|c|c|l|}
	\hline
	\textbf{Parameter} & \textbf{Units} & \textbf{Description} \\  \hline
	ECS & $^{\circ}$C & Equilibrium climate sensitivity \\ \hline
	$\kappa_{DOECLIM}$ & cm$^2$ s$^{-1}$ & Ocean vertical diffusivity \\ \hline
	$\alpha_{DOECLIM}$ & $-$ & Aerosol scaling factor\\ \hline
	$T_0$ & K & Global mean surface temperature initial condition \\ \hline
	$\sigma_T$ & K & Global mean surface temperature AR1 innovation standard deviation \\ \hline
	$\rho_T$ & $-$ & Global mean surface temperature autocorrelation\\ \hline
	$H_0$ & $10^{22}$J & Ocean heat uptake initial condition\\ \hline
	$\sigma_H$ & $10^{22}$J & Ocean heat uptake AR1 innovation standard deviation \\ \hline
	$\rho_H$ & $-$ & Ocean heat uptake autocorrelation\\ \hline
	$\beta_0$ & m yr$^{-1}$ & Initial GSIC mass balance sensitivity \\ \hline
	$V_{0, GSIC}$ & m & Initial GSIC volume\\ \hline
	$n$ & $-$ & GSIC exponent for area-volume scaling\\ \hline
	$G_{s, 0}$ & m & Sea-level rise from GSIC in the first model year \\ \hline
	$\sigma_{GSIC}$ & m & GSIC AR1 innovation standard deviation \\ \hline
	$\rho_{GSIC}$ & $-$ & GSIC autocorrelation\\ \hline
	$\alpha_{TE}$ & kg m$^{-3}$ K$^{-1}$ & Thermal expansion coefficient\\ \hline
	$a_{SIMPLE}$ & m K$^{-1}$ & Temperature sensitivity of equilibrium GIS volume\\ \hline
	$b_{SIMPLE}$ & m & Equilibrium GIS volume for temperature anomaly of 0$^{\circ}$C\\ \hline
	$\alpha_{SIMPLE}$ & yr$^{-1}$ K$^{-1}$ & Temperature sensitivity of GIS exponential decay rate\\ \hline
	$\beta_{SIMPLE}$ & yr$^{-1}$ & GIS exponential decay rate for temperature anomaly of 0$^{\circ}$C \\ \hline
	$V_{0, SIMPLE}$ & m & Initial GIS volume\\ \hline
	$\sigma_{GIS}$ & m  & GIS AR1 innovation standard deviation\\ \hline
	$\rho_{GIS}$ & $-$ & GIS autocorrelation\\ \hline
	$a_{ANTO}$ & $^{\circ}$C $^{\circ}$C$^{-1}$ & Sensitivity of Antarctic ocean temperature to surface temperature    \\ \hline
	$b_{ANTO}$ & $^{\circ}$C & Antarctic ocean temperature for surface temperature anomaly of 0$^{\circ}$C\\ \hline
	$\gamma$ & $-$ & Power for the relation of ice flow speed to water depth  \\ \hline
	$\alpha_{DAIS}$ & $-$ & Partition parameter for effect of ocean subsurface temperature on ice flux \\ \hline
	$\mu$ & m$^{1/2}$ & Profile parameter for parabolic Antarctic ice sheet surface (related to ice stress) \\ \hline
	$\nu$ & m$^{-1/2}$ yr$^{-1/2}$ & Proportionality constant relating runoff decrease with height to precipitation \\ \hline
	$P_0$ & m yr$^{-1}$ & Antarctic annual precipitation for Antarctic surface temperature of 0$^{\circ}$C  \\ \hline
	$\kappa_{DAIS}$ & $^{\circ}$C$^{-1}$ & Coefficient for exponential dependency of precipitation on Antarctic temperature   \\ \hline
	$f_0$ & m yr$^{-1}$ & Proportionality constant for ice flow at grounding line \\ \hline
	$h_0$ & m & Height of runoff line at Antarctic surface temperature of 0$^{\circ}$C \\ \hline
	$C$ & m $^{\circ}$C$^{-1}$ & Sensitivity of height of runoff line \\ \hline
	$b_0$ & m & Undisturbed bed height at the Antarctic continent center \\ \hline
	$slope$ & $-$ & Slope of ice sheet bed before loading \\ \hline
	$\lambda$ & mm yr$^{-1}$ & Rate of fast dynamical disintegration of the Antarctic ice sheet \\ \hline
	$T_{crit}$ & $^{\circ}$C &  Temperature associated with onset of fast dynamical disintegration of the Antarctic ice sheet\\ \hline
	\end{tabular}
\end{table}

\clearpage

\begin{table}[t]
    \caption{Best 10 random forest hyperparameter sets for RCP2.6 and RCP8.5 in 2050 (max\_features hyperparameter fixed at `sqrt')}
    \label{2050_hyperparameter_table}
    \begin{tabular}{ccccccc}
    \hline
    & & & & & & {\footnotesize \textbf{Mean Cross-}} \\
    & & & & & & {\footnotesize \textbf{Validation}} \\
	& {\footnotesize \textbf{Rank}} & {\footnotesize \textbf{max\_depth}} & {\footnotesize \textbf{min\_samples\_leaf}} & {\footnotesize \textbf{min\_samples\_split}} & {\footnotesize \textbf{n\_estimators}} & {\footnotesize \textbf{Accuracy}} \\ \hline
    RCP2.6 2050 & 1 & 18 & 4 & 7 & 750 & 0.92075 \\
    & 2 & 12 & 4 & 7 & 250 & 0.920375   \\
    & 3 & 18 & 4 & 7 & 1000 & 0.920125  \\
    & 4 & 14 & 4 & 10 & 500 & 0.92    \\
    & 4 & 16 & 4 & 10 & 250 & 0.92  \\
    & 4 & 18 & 4 & 7 & 250 & 0.92 \\
    & 7 & 14 & 4 & 7 & 250 & 0.919875  \\
    & 7 & 14 & 4 & 7 & 500 & 0.919875  \\
    & 7 & 14 & 4 & 7 & 750 & 0.919875  \\
    & 10 & 18 & 4 & 13 & 500 & 0.91975  \\
    & 10 & 18 & 4 & 10 & 1000 & 0.91975 \\
    & 10 & 18 & 4 & 10 & 750 & 0.91975 \\
    \hline
    RCP8.5 2050 & 1 & 16 & 4 & 10 & 250 & 0.9265  \\
    & 2 & 16 & 4 & 7 & 500 & 0.925625    \\
    & 2 & 14 & 4 & 7 & 1000 & 0.925625    \\\
    & 2 & 12 & 4 & 7 & 250 & 0.925625    \\
    & 5 & 16 & 4 & 13 & 250 & 0.9255  \\
    & 5 & 14 & 4 & 7 & 500 & 0.9255  \\
    & 7 & 16 & 4 & 16 & 250 & 0.925375    \\
    & 7 & 16 & 4 & 7 & 1000 & 0.925375    \\
    & 9 & 18 & 4 & 7 & 750 & 0.92525 \\
    & 10 & 14 & 4 & 10 & 250 & 0.925125    \\
    & 10 & 18 & 4 & 7 & 500 & 0.925125    \\
    \end{tabular}
\end{table}

\clearpage

\begin{table}[t]
    \caption{Best 10 random forest hyperparameter sets for RCP2.6 and RCP8.5 in 2075 (max\_features hyperparameter fixed at `sqrt')}
    \label{2075_hyperparameter_table}
    \begin{tabular}{ccccccc}
    \hline
    & & & & & & {\footnotesize \textbf{Mean Cross-}} \\
    & & & & & & {\footnotesize \textbf{Validation}} \\
	& {\footnotesize \textbf{Rank}} & {\footnotesize \textbf{max\_depth}} & {\footnotesize \textbf{min\_samples\_leaf}} & {\footnotesize \textbf{min\_samples\_split}} & {\footnotesize \textbf{n\_estimators}} & {\footnotesize \textbf{Accuracy}} \\ \hline
    RCP2.6 2075 & 1 & 16 & 4 & 7 & 500 & 0.942 \\
    & 2 & 14 & 4 & 7 & 750 & 0.94175    \\
    & 3 & 14 & 4 & 10 & 500 & 0.941625 \\
    & 4 & 16 & 4 & 7 & 750 & 0.9415 \\
    & 4 & 18 & 4 & 10 & 750 & 0.9415    \\
    & 4 & 14 & 4 & 7 & 1000 & 0.9415    \\
    & 4 & 16 & 4 & 10 & 1000 & 0.9415   \\
    & 4 & 14 & 4 & 16 & 500 & 0.9415    \\
    & 9 & 18 & 4 & 10 & 250 & 0.941375  \\
    & 9 & 18 & 4 & 7 & 750 & 0.941375   \\
    & 9 & 12 & 4 & 7 & 250 & 0.941375   \\
    & 9 & 12 & 4 & 7 & 500 & 0.941375   \\
    \hline
    RCP8.5 2075 & 1 & 14 & 4 & 7 & 750 & 0.920625 \\
    & 2 & 16 & 4 & 7 & 500 & 0.9205  \\
    & 3 & 16 & 4 & 7 & 750 & 0.920375 \\
    & 4 & 12 & 4 & 7 & 500 & 0.92025 \\
    & 5 & 18 & 4 & 7 & 500 & 0.92    \\
    & 6 & 12 & 4 & 10 & 750 & 0.919875    \\
    & 6 & 16 & 4 & 7 & 1000 & 0.919875    \\
    & 8 & 16 & 4 & 10 & 1000 & 0.91975 \\
    & 8 & 16 & 4 & 13 & 250 & 0.91975 \\
    & 10 & 18 & 4 & 13 & 500 & 0.919625    \\
    & 10 & 18 & 4 & 10 & 1000 & 0.919625    \\
    & 10 & 18 & 4 & 7 & 750 & 0.919625    \\
    & 10 & 12 & 4 & 13 & 750 & 0.919625    \\
    & 10 & 14 & 4 & 10 & 750 & 0.919625    \\
    & 10 & 14 & 4 & 7 & 1000 & 0.919625    \\
    \end{tabular}
\end{table}

\clearpage

\begin{table}
    \caption{Best 10 random forest hyperparameter sets for RCP2.6 and RCP8.5 in 2100 (max\_features hyperparameter fixed at `sqrt')}
    \label{2100_hyperparameter_table}
    \begin{tabular}{ccccccc}
    \hline
    & & & & & & {\footnotesize \textbf{Mean Cross-}} \\
    & & & & & & {\footnotesize \textbf{Validation}} \\
	& {\footnotesize \textbf{Rank}} & {\footnotesize \textbf{max\_depth}} & {\footnotesize \textbf{min\_samples\_leaf}} & {\footnotesize \textbf{min\_samples\_split}} & {\footnotesize \textbf{n\_estimators}} & {\footnotesize \textbf{Accuracy}} \\ \hline
    RCP2.6 2100 & 1 & 12 & 4 & 7 & 1000 & 0.9515  \\
    & 2 & 16 & 4 & 7 & 500 & 0.951375    \\
    & 3 & 16 & 4 & 7 & 1000 & 0.95125 \\
    & 4 & 18 & 4 & 13 & 500 & 0.951125    \\
    & 4 & 14 & 4 & 10 & 250 & 0.951125  \\
    & 6 & 12 & 4 & 7 & 250 & 0.951   \\
    & 6 & 14 & 4 & 7 & 250 & 0.951   \\
    & 8 & 14 & 4 & 7 & 750 & 0.950875    \\
    & 8 & 18 & 4 & 10 & 500 & 0.950875    \\
    & 10 & 18 & 4 & 7 & 750 & 0.95075 \\
    & 10 & 18 & 4 & 13 & 1000 & 0.95075 \\
    & 10 & 14 & 4 & 10 & 750 & 0.95075 \\
    & 10 & 14 & 4 & 7 & 1000 & 0.95075 \\
    & 10 & 12 & 4 & 7 & 750 & 0.95075 \\
    \hline
    RCP8.5 2100 & 1 & 18 & 4 & 7 & 250 & 0.924625    \\
    & 2 & 16 & 4 & 7 & 250 & 0.923875    \\
    & 3 & 16 & 4 & 10 & 250 & 0.92375 \\
    & 4 & 14 & 4 & 7 & 1000 & 0.923375    \\
    & 4 & 16 & 4 & 10 & 500 & 0.923375    \\
    & 4 & 18 & 4 & 10 & 1000 & 0.923375    \\
    & 7 & 18 & 4 & 7 & 500 & 0.92325 \\
    & 7 & 16 & 4 & 13 & 250 & 0.92325 \\
    & 7 & 18 & 4 & 13 & 250 & 0.92325 \\
    & 10 & 16 & 4 & 7 & 500 & 0.923125    \\
    & 10 & 18 & 4 & 10 & 500 & 0.923125    \\
    & 10 & 18 & 4 & 10 & 750 & 0.923125    \\
    \end{tabular}
\end{table}

\clearpage

\begin{table}[t]
    \centering
    \caption{Random forests' accuracy on the training and testing subsets}
    \label{forest_accuracy}
    \begin{tabular}{c|cc|cc}
        & \multicolumn{2}{c}{\textbf{RCP2.6}} & \multicolumn{2}{c}{\textbf{RCP8.5}} \\
        \textbf{Year} & \textbf{Training Accuracy} & \textbf{Test Accuracy} & \textbf{Training Accuracy} & \textbf{Test Accuracy}  \\ \hline
        2020 & 0.955 & 0.9095 & 0.953875 & 0.8905 \\
        2025 & 0.963 & 0.906 & 0.96175 & 0.902\\
        2030 & 0.966875 & 0.9095 & 0.966 & 0.9055\\ 
        2035 & 0.971 & 0.902 & 0.971125 & 0.9115\\ 
        2040 & 0.976875 & 0.904 & 0.976625 & 0.922 \\ 
        2045 & 0.97825 & 0.9175 & 0.98375 & 0.9315 \\ 
        2050 & 0.982 & 0.935 & 0.9895 & 0.9245 \\ 
        2055 & 0.983 & 0.9385 & 0.989875 & 0.9285 \\ 
        2060 & 0.981875 & 0.93 & 0.988875 & 0.915 \\ 
        2065 & 0.98675 & 0.9375 & 0.990875 & 0.925 \\ 
        2070 & 0.986 & 0.9405 & 0.989125 & 0.9195 \\ 
        2075 & 0.98725 & 0.943 & 0.98925 & 0.9245 \\ 
        2080 & 0.9875 & 0.938 & 0.98925 & 0.9255 \\
        2085 & 0.989375 & 0.947 & 0.98775 & 0.9185 \\ 
        2090 & 0.99 & 0.954 & 0.9905 & 0.926 \\ 
        2095 & 0.990625 & 0.945 & 0.99 & 0.9295 \\ 
        2100 & 0.99125 & 0.9485 & 0.988875 & 0.9285 \\ 
        2105 & 0.991 & 0.952 & 0.990875 & 0.93 \\ 
        2110 & 0.991375 & 0.9525 & 0.988875 & 0.9175 \\ 
        2115 & 0.991375 & 0.964 & 0.990125 & 0.9275 \\ 
        2120 & 0.99125 & 0.948 & 0.989875 & 0.9355 \\ 
        2125 & 0.9915 & 0.9545 & 0.988875 & 0.932 \\ 
        2130 & 0.9925 & 0.9515 & 0.989875 & 0.9255 \\ 
        2135 & 0.9925 & 0.953 & 0.990875 & 0.938 \\ 
        2140 & 0.993125 & 0.9535 & 0.99025 & 0.9385 \\ 
        2145 & 0.991375 & 0.9545 & 0.990875 & 0.938 \\ 
        2150 & 0.9925 & 0.949 & 0.99175 & 0.937 \\
    \end{tabular}
\end{table}

\clearpage

\begin{table}[t]
    \centering
    \caption{Quartile descriptions of the distributions of the maximum equilibrium climate sensitivity (ECS) split value from each decision tree in the fitted forests.  Q1 denotes Quartile 1, which is the 25th percentile of the data.  The median is the 50th percentile of the data.  Q3 denotes Quartile 3, which is the 75th percentile of the data.  IQR stands for the interquartile range, which is calculated as Q3$-$Q1.}
    \label{max_ECS_split_quartiles}
    \begin{tabular}{c|cccc|cccc}
        & \multicolumn{4}{c}{\textbf{RCP2.6}} & \multicolumn{4}{c}{\textbf{RCP8.5}} \\
        \textbf{Year} & \textbf{Q1} & \textbf{Median} & \textbf{Q3} & \textbf{IQR} & \textbf{Q1} & \textbf{Median} & \textbf{Q3} & \textbf{IQR}  \\ \hline
        2020 & 5.2730 & 5.9075 & 6.8610 & 1.5880 & 5.2815 & 6.2893 & 7.1646 & 1.8831\\
        2030 & 5.6522 & 6.2781 & 7.0024 & 1.3502 & 5.8065 & 6.3252 & 7.0982 & 1.2918\\
        2040 & 5.9763 & 6.4340 & 7.0379 & 1.0617 & 6.0098 & 6.5179 & 7.1364 & 1.1266\\
        2050 & 5.8078 & 6.3209 & 6.7734 & 0.9656 & 6.1071 & 6.5088 & 6.9855 & 0.8784        \\
        2060 & 5.8036 & 6.2169 & 6.9536 & 1.1500 & 5.9752 & 6.4841 & 7.0021 & 1.0269        \\
        2070 & 6.0922 & 6.4180 & 6.6895 & 0.5972 & 5.9310 & 6.4693 & 7.2775 & 1.3465        \\
        2080 & 6.2783 & 6.4861 & 6.9545 & 0.6762 & 5.8077 & 6.2986 & 6.9806 & 1.1728        \\
        2090 & 6.1806 & 6.4130 & 6.7686 & 0.5880 & 5.9089 & 6.3977 & 7.3010 & 1.3922        \\
        2100 & 6.2350 & 6.4344 & 6.9115 & 0.6765 & 5.9409 & 6.5145 & 7.3476 & 1.4068        \\
        2110 & 6.2494 & 6.4822 & 6.9536 & 0.7041 & 5.8222 & 6.3857 & 7.3446 & 1.5223        \\
        2120 & 6.1750 & 6.3773 & 6.7036 & 0.5286 & 5.9547 & 6.6390 & 7.3376 & 1.3829        \\
        2130 & 6.2525 & 6.5197 & 7.0188 & 0.7662 & 5.8576 & 6.3889 & 7.0532 & 1.1955        \\
        2140 & 6.1830 & 6.4520 & 6.9820 & 0.7991 & 5.9409 & 6.4173 & 7.1532 & 1.2123        \\
        2150 & 6.3337 & 6.5538 & 7.0190 & 0.6853 & 5.9857 & 6.4986 & 7.3002 & 1.3144 \\
    \end{tabular}
\end{table}

\clearpage

\begin{figure*}[t]
    \includegraphics[width=12cm]{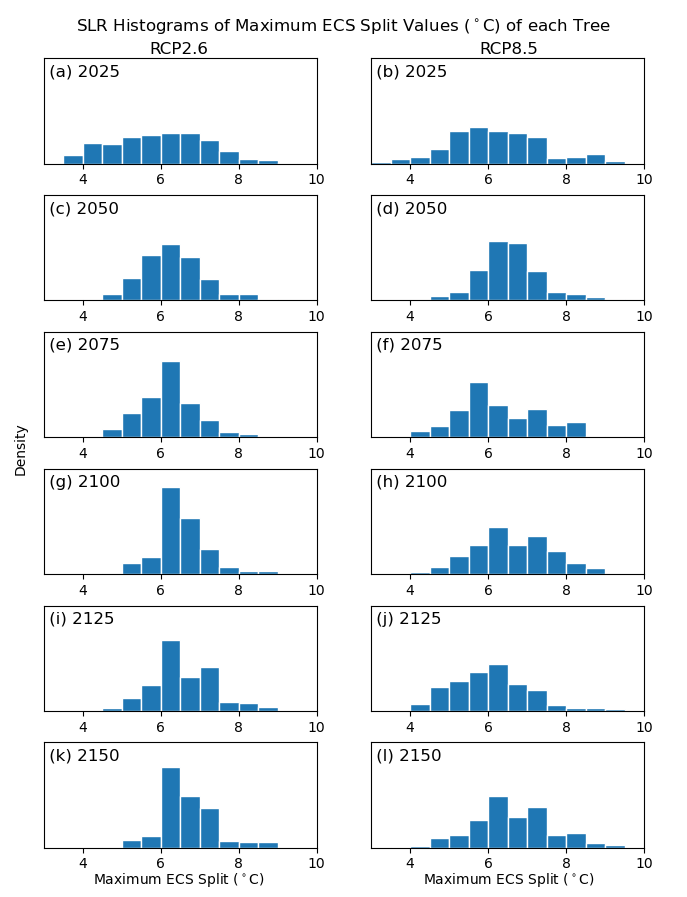}
    \caption{Distributions of the maximum equilibrium climate sensitivity (ECS) split value from each decision tree in the fitted random forests.  The left column of plots depicts the maximum ECS split distributions in the RCP2.6 forests, and the right column depicts the maximum ECS split distributions in the RCP8.5 forests.}
    \label{max_ECS_histogram}
\end{figure*}

\clearpage

\begin{figure*}[t]
    \includegraphics[width=12cm]{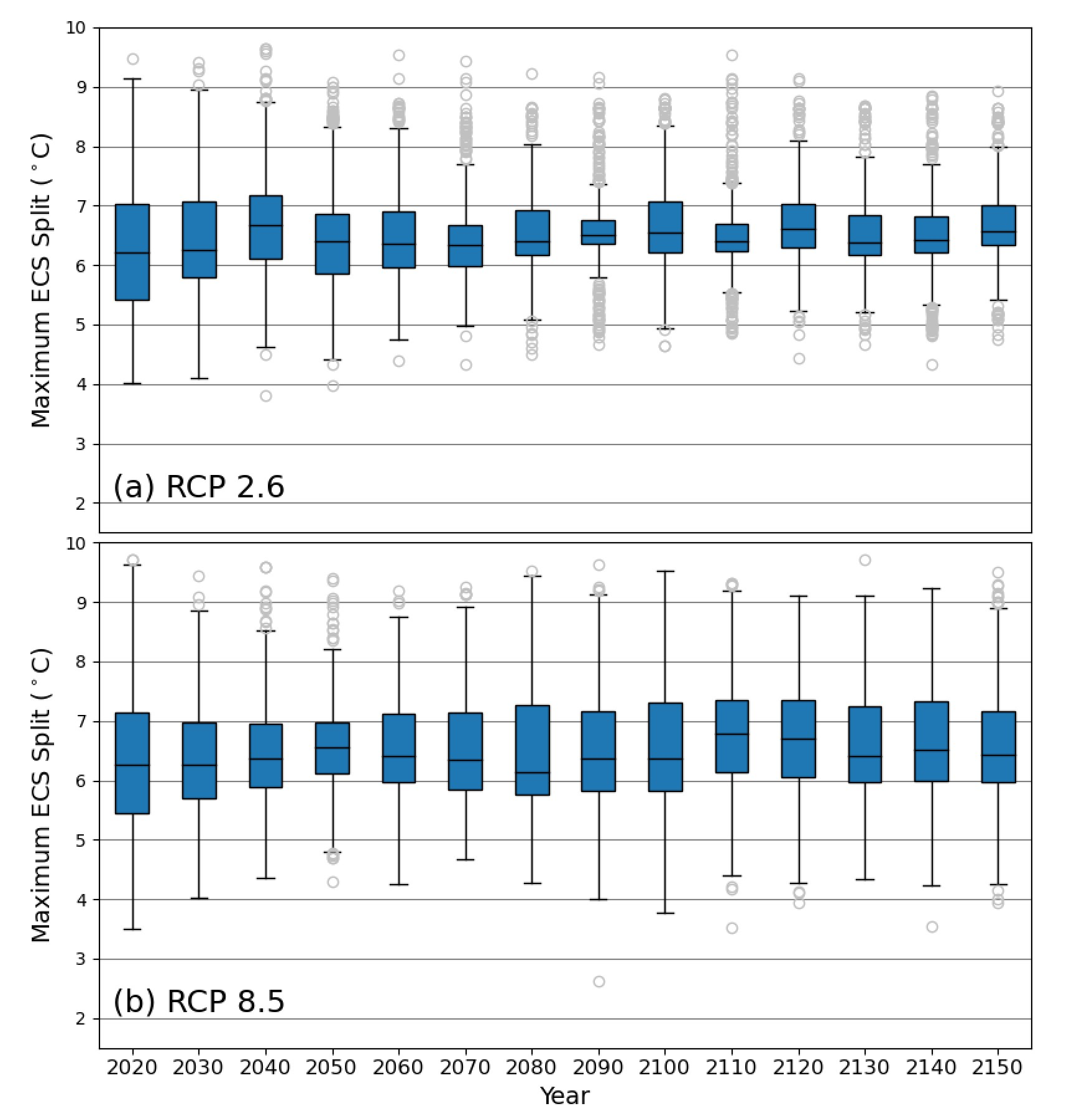}
    \caption{Distributions of the maximum equilibrium climate sensitivity (ECS) split value from each decision tree in the random forests fitted using the 80th percentile to classify the global mean sea level (GMSL) data.  (a) depicts the maximum ECS split distributions in the RCP2.6 forests, and (b) depicts the maximum ECS split distributions in the RCP8.5 forests.  The outliers are the data points less than $\text{Q1} - 1.5* \text{IQR}$ or greater than $\text{Q3} + 1.5* \text{IQR}$, where Q1 is Quartile 1, Q3 is Quartile 3, and IQR is the interquartile range (Q3$-$Q1).  The blue boxes show the IQR, and the line within the IQR is the median.}
    \label{80th_percentile_max_ECS_boxplot}
\end{figure*}

\clearpage

\noappendix       




\appendixfigures  

\appendixtables   


\authorcontribution{TW designed the research. AH performed research and created the figures. AH prepared the first draft of the manuscript, and all authors contributed to the final version of the manuscript.} 

\competinginterests{The authors declare no competing interests.} 


\begin{acknowledgements}
The authors gratefully acknowledge support from Rochester Institute of Technology’s Sponsored Research Services Seed Grant program. The authors thank Ben Vega-Westhoff, Vivek Srikrishnan, Frank Errickson, and Radley Powers for valuable inputs.
\end{acknowledgements}




\bibliographystyle{copernicus}
\bibliography{manuscript.bib}

\begin{thebibliography}{38}
\providecommand{\natexlab}[1]{#1}
\providecommand{\url}[1]{{\tt #1}}
\providecommand{\urlprefix}{URL }
\expandafter\ifx\csname urlstyle\endcsname\relax
  \providecommand{\doi}[1]{https://doi.org/\discretionary{}{}{}#1}\else
  \providecommand{\doi}{https://doi.org/\discretionary{}{}{}\begingroup
  \urlstyle{rm}\Url}\fi

\bibitem[{Bakker et~al.(2016)Bakker, Applegate, and Keller}]{SIMPLE}
Bakker, A.~M., Applegate, P.~J., and Keller, K.: A simple, physically motivated
  model of sea-level contributions from the Greenland ice sheet in response to
  temperature changes, Environmental Modelling \& Software, 83, 27--35,
  \doi{https://doi.org/10.1016/j.envsoft.2016.05.003}, 2016.

\bibitem[{Bakker et~al.(2017)Bakker, Wong, Ruckert, and Keller}]{ANTO}
Bakker, A. M.~R., Wong, T.~E., Ruckert, K.~L., and Keller, K.: Sea-level
  projections representing the deeply uncertain contribution of the West
  Antarctic ice sheet, Scientific Reports, 7, 3880,
  \doi{10.1038/s41598-017-04134-5}, 2017.

\bibitem[{Church et~al.(2013)Church, Clark, Cazenave, Gregory, Jevrejeva,
  Levermann, Merrifield, Milne, Nerem, Nunn, Payne, Pfeffer, Stammer, and
  Unnikrishnan}]{land_water_storage}
Church, J., Clark, P., Cazenave, A., Gregory, J., Jevrejeva, S., Levermann, A.,
  Merrifield, M., Milne, G., Nerem, R., Nunn, P., Payne, A., Pfeffer, W.,
  Stammer, D., and Unnikrishnan, A.: Sea Level Change, book section~13, p.
  1137–1216, Cambridge University Press, Cambridge, United Kingdom and New
  York, NY, USA, \doi{10.1017/CBO9781107415324.026}, 2013.

\bibitem[{Dayan et~al.(2021)Dayan, Le~Cozannet, Speich, and
  Thi{\'e}blemont}]{risk_aversion}
Dayan, H., Le~Cozannet, G., Speich, S., and Thi{\'e}blemont, R.: High-End
  Scenarios of Sea-Level Rise for Coastal Risk-Averse Stakeholders, Frontiers
  in Marine Science, 8, 514, \doi{10.3389/fmars.2021.569992}, 2021.

\bibitem[{DeConto et~al.(2021)DeConto, Pollard, Alley, Velicogna, Gasson,
  Gomez, Sadai, Condron, Gilford, Ashe, Kopp, Li, and
  Dutton}]{AIS_disintegration_2}
DeConto, R.~M., Pollard, D., Alley, R.~B., Velicogna, I., Gasson, E., Gomez,
  N., Sadai, S., Condron, A., Gilford, D.~M., Ashe, E.~L., Kopp, R.~E., Li, D.,
  and Dutton, A.: The Paris Climate Agreement and future sea-level rise from
  Antarctica, Nature, 593, 83--89, \doi{10.1038/s41586-021-03427-0}, 2021.

\bibitem[{{Exec. Order No. 14008}(2021)}]{climate_gov_doc_1}
{Exec. Order No. 14008}: Tackling the Climate Crisis at Home and Abroad, 86
  F.R. 7619, 7619--7633,
  \urlprefix\url{https://www.federalregister.gov/documents/2021/02/01/2021-02177/tackling-the-climate-crisis-at-home-and-abroad},
  2021.

\bibitem[{FAO(2014)}]{coast_stats}
FAO: Global Blue Growth Initiative and Small Island Developing States (SIDS),
  Food and Agriculture Organization of the United Nations (FAO),
  \urlprefix\url{http://www.fao.org/documents/card/en/c/c8aeb23f-f794-410e-804f-2aa82140d34a/},
  2014.

\bibitem[{Fischbach et~al.(2012)Fischbach, Johnson, Ortiz, Bryant, Hoover, and
  Ostwald}]{CLARA}
Fischbach, J.~R., Johnson, D.~R., Ortiz, D.~S., Bryant, B.~P., Hoover, M., and
  Ostwald, J.: Coastal Louisiana Risk Assessment Model: Technical Description
  and 2012 Coastal Master Plan Analysis Results, RAND Corporation, Santa
  Monica, CA, 2012.

\bibitem[{Flato et~al.(2013)Flato, Marotzke, Abiodun, Braconnot, Chou, Collins,
  Cox, Driouech, Emori, Eyring, Forest, Gleckler, Guilyardi, Jakob, Kattsov,
  Reason, and Rummukainen}]{climate_model_uncertainties_1}
Flato, G., Marotzke, J., Abiodun, B., Braconnot, P., Chou, S.~C., Collins, W.,
  Cox, P., Driouech, F., Emori, S., Eyring, V., Forest, C., Gleckler, P.,
  Guilyardi, E., Jakob, C., Kattsov, V., Reason, C., and Rummukainen, M.:
  Evaluation of climate models, book section~9, pp. 741--882, Cambridge
  University Press, Cambridge, United Kingdom and New York, NY, USA,
  \doi{10.1017/CBO9781107415324.020}, 2013.

\bibitem[{Ga{\'a}l et~al.(2012)Ga{\'a}l, Moriondo, and Bindi}]{similar_work_2}
Ga{\'a}l, M., Moriondo, M., and Bindi, M.: Modelling the impact of climate
  change on the Hungarian wine regions using Random Forest, Applied Ecology and
  Environmental Research, 10, 121--140, \doi{10.15666/aeer/1002_121140}, 2012.

\bibitem[{Giorgi(2019)}]{climate_model_uncertainties_2}
Giorgi, F.: Thirty Years of Regional Climate Modeling: Where Are We and Where
  Are We Going next?, Journal of Geophysical Research: Atmospheres, 124,
  5696--5723, \doi{https://doi.org/10.1029/2018JD030094}, 2019.

\bibitem[{Grinsted et~al.(2010)Grinsted, Moore, and Jevrejeva}]{TE_1}
Grinsted, A., Moore, J., and Jevrejeva, S.: Reconstructing sea level from paleo
  and projected temperatures 200 to 2100 AD, Climate Dynamics, 34, 461--472,
  \doi{10.1007/s00382-008-0507-2}, paper id:: 10.1007/s00382-008-0507-2, 2010.

\bibitem[{Haasnoot et~al.(2013)Haasnoot, Kwakkel, Walker, and {ter
  Maat}}]{dynamic_adaptive_pathways_1}
Haasnoot, M., Kwakkel, J.~H., Walker, W.~E., and {ter Maat}, J.: Dynamic
  adaptive policy pathways: A method for crafting robust decisions for a deeply
  uncertain world, Global Environmental Change, 23, 485--498,
  \doi{https://doi.org/10.1016/j.gloenvcha.2012.12.006}, 2013.

\bibitem[{Hartin et~al.(2015)Hartin, Patel, Schwarber, Link, and
  Bond-Lamberty}]{hector}
Hartin, C.~A., Patel, P., Schwarber, A., Link, R.~P., and Bond-Lamberty, B.~P.:
  A simple object-oriented and open-source model for scientific and policy
  analyses of the global climate system – Hector v1.0, Geoscientific Model
  Development, 8, 939--955, \doi{10.5194/gmd-8-939-2015}, 2015.

\bibitem[{Hinkel et~al.(2014)Hinkel, Lincke, Vafeidis, Perrette, Nicholls, Tol,
  Marzeion, Fettweis, Ionescu, and Levermann}]{climate_gov_doc_3}
Hinkel, J., Lincke, D., Vafeidis, A.~T., Perrette, M., Nicholls, R.~J., Tol, R.
  S.~J., Marzeion, B., Fettweis, X., Ionescu, C., and Levermann, A.: Coastal
  flood damage and adaptation costs under 21st century sea-level rise,
  Proceedings of the National Academy of Sciences, 111, 3292--3297,
  \doi{10.1073/pnas.1222469111}, 2014.

\bibitem[{James et~al.(2013)James, Witten, Hastie, and
  Tibshirani}]{ISL_textbook}
James, G., Witten, D., Hastie, T., and Tibshirani, R.: An Introduction to
  Statistical Learning: With Applications in R, Springer,
  \doi{https://doi.org/10.1007/978-1-4614-7138-7}, 2013.

\bibitem[{Jelesnianski et~al.(1992)Jelesnianski, Chen, Shaffer, sms, and
  Service}]{SLOSH}
Jelesnianski, C., Chen, J., Shaffer, W., sms, . O.~., and Service, U. S. N.~W.:
  SLOSH: Sea, Lake, and Overland Surges from Hurricanes, NOAA technical report
  NWS, U.S. Department of Commerce, National Oceanic and Atmospheric
  Administration, National Weather Service,
  \urlprefix\url{https://books.google.com/books?id=Wdg8mQfzkVcC}, 1992.

\bibitem[{Kopp et~al.(2017)Kopp, DeConto, Bader, Hay, Horton, Kulp,
  Oppenheimer, Pollard, and Strauss}]{AIS_disintegration_1}
Kopp, R.~E., DeConto, R.~M., Bader, D.~A., Hay, C.~C., Horton, R.~M., Kulp, S.,
  Oppenheimer, M., Pollard, D., and Strauss, B.~H.: Evolving Understanding of
  Antarctic Ice-Sheet Physics and Ambiguity in Probabilistic Sea-Level
  Projections, Earth's Future, 5, 1217--1233,
  \doi{https://doi.org/10.1002/2017EF000663}, 2017.

\bibitem[{Kriegler(2005)}]{DOECLIM}
Kriegler, E.: Imprecise probability analysis for integrated assessment of
  climate change, doctoralthesis, Universit{\"a}t Potsdam,
  \urlprefix\url{http://opus.kobv.de/ubp/volltexte/2005/561/}, 2005.

\bibitem[{{Le Cozannet} et~al.(2015){Le Cozannet}, Rohmer, Cazenave, Idier,
  {van de Wal}, {de Winter}, Pedreros, Balouin, Vinchon, and
  Oliveros}]{climate_gov_doc_4}
{Le Cozannet}, G., Rohmer, J., Cazenave, A., Idier, D., {van de Wal}, R., {de
  Winter}, R., Pedreros, R., Balouin, Y., Vinchon, C., and Oliveros, C.:
  Evaluating uncertainties of future marine flooding occurrence as sea-level
  rises, Environmental Modelling \& Software, 73, 44--56,
  \doi{https://doi.org/10.1016/j.envsoft.2015.07.021}, 2015.

\bibitem[{Meinshausen et~al.(2011)Meinshausen, Raper, and Wigley}]{GIC-MAGICC}
Meinshausen, M., Raper, S. C.~B., and Wigley, T. M.~L.: Emulating coupled
  atmosphere-ocean and carbon cycle models with a simpler model, MAGICC6 –
  Part 1: Model description and calibration, Atmospheric Chemistry and Physics,
  11, 1417--1456, \doi{10.5194/acp-11-1417-2011}, 2011.

\bibitem[{Mengel et~al.(2016)Mengel, Levermann, Frieler, Robinson, Marzeion,
  and Winkelmann}]{TE_2}
Mengel, M., Levermann, A., Frieler, K., Robinson, A., Marzeion, B., and
  Winkelmann, R.: Future sea level rise constrained by observations and
  long-term commitment, Proceedings of the National Academy of Sciences, 113,
  2597--2602, \doi{10.1073/pnas.1500515113}, 2016.

\bibitem[{Moss et~al.(2010)Moss, Edmonds, Hibbard, Manning, Rose, van Vuuren,
  Carter, Emori, Kainuma, Kram, Meehl, Mitchell, Nakicenovic, Riahi, Smith,
  Stouffer, Thomson, Weyant, and Wilbanks}]{RCP}
Moss, R.~H., Edmonds, J.~A., Hibbard, K.~A., Manning, M.~R., Rose, S.~K., van
  Vuuren, D.~P., Carter, T.~R., Emori, S., Kainuma, M., Kram, T., Meehl, G.~A.,
  Mitchell, J. F.~B., Nakicenovic, N., Riahi, K., Smith, S.~J., Stouffer,
  R.~J., Thomson, A.~M., Weyant, J.~P., and Wilbanks, T.~J.: The next
  generation of scenarios for climate change research and assessment, Nature,
  463, 747--756, \doi{10.1038/nature08823}, 2010.

\bibitem[{Nauels et~al.(2017)Nauels, Rogelj, Schleussner, Meinshausen, and
  Mengel}]{AIS_disintegration_4}
Nauels, A., Rogelj, J., Schleussner, C.-F., Meinshausen, M., and Mengel, M.:
  Linking sea level rise and socioeconomic indicators under the Shared
  Socioeconomic Pathways, Environmental Research Letters, 12, 114\,002,
  \doi{10.1088/1748-9326/aa92b6}, 2017.

\bibitem[{{New Orleans Health Department}(2018)}]{climate_gov_doc_2}
{New Orleans Health Department}: Climate Change \& Health Report,
  \urlprefix\url{https://www.nola.gov/getattachment/Health/Climate-Change-(1)/Planning-Tools-and-Data/Climate-Change-and-Health-Report-2018-Final.pdf/},
  2018.

\bibitem[{Pedregosa et~al.(2011)Pedregosa, Varoquaux, Gramfort, Michel,
  Thirion, Grisel, Blondel, Prettenhofer, Weiss, Dubourg, Vanderplas, Passos,
  Cournapeau, Brucher, Perrot, and Duchesnay}]{scikit-learn}
Pedregosa, F., Varoquaux, G., Gramfort, A., Michel, V., Thirion, B., Grisel,
  O., Blondel, M., Prettenhofer, P., Weiss, R., Dubourg, V., Vanderplas, J.,
  Passos, A., Cournapeau, D., Brucher, M., Perrot, M., and Duchesnay, E.:
  Scikit-learn: Machine Learning in {P}ython, Journal of Machine Learning
  Research, 12, 2825--2830, 2011.

\bibitem[{{Python 3.7.4}(2019)}]{python}
{Python 3.7.4}: Python Language Reference, Python Software Foundation,
  \urlprefix\url{https://www.python.org/}, 2019.

\bibitem[{{R Core Team}(2019)}]{R}
{R Core Team}: R: A Language and Environment for Statistical Computing, R
  Foundation for Statistical Computing, Vienna, Austria,
  \urlprefix\url{https://www.R-project.org/}, 2019.

\bibitem[{Rohmer et~al.(2021)Rohmer, Lincke, Hinkel, Le~Cozannet, Lambert, and
  Vafeidis}]{similar_work_1}
Rohmer, J., Lincke, D., Hinkel, J., Le~Cozannet, G., Lambert, E., and Vafeidis,
  A.~T.: Unravelling the Importance of Uncertainties in Global-Scale Coastal
  Flood Risk Assessments under Sea Level Rise, Water, 13,
  \doi{10.3390/w13060774}, 2021.

\bibitem[{Ruckert et~al.(2019)Ruckert, Srikrishnan, and
  Keller}]{dynamic_adaptive_pathways_3}
Ruckert, K.~L., Srikrishnan, V., and Keller, K.: Characterizing the deep
  uncertainties surrounding coastal flood hazard projections: A case study for
  Norfolk, VA, Scientific Reports, 9, 11\,373,
  \doi{10.1038/s41598-019-47587-6}, 2019.

\bibitem[{Shaffer(2014)}]{DAIS}
Shaffer, G.: Formulation, calibration and validation of the DAIS model (version
  1), a simple Antarctic ice sheet model sensitive to variations of sea level
  and ocean subsurface temperature, Geoscientific Model Development, 7,
  1803--1818, \doi{10.5194/gmd-7-1803-2014}, 2014.

\bibitem[{Vega-Westhoff(2019)}]{vega_westhoff_data}
Vega-Westhoff, B.: {Updated MCMC chains and subsamples for Hector calibration
  paper}, \doi{10.5281/zenodo.3236413}, 2019.

\bibitem[{Vega-Westhoff et~al.(2019)Vega-Westhoff, Sriver, Hartin, Wong, and
  Keller}]{supplemental_parameter_table}
Vega-Westhoff, B., Sriver, R.~L., Hartin, C.~A., Wong, T.~E., and Keller, K.:
  Impacts of Observational Constraints Related to Sea Level on Estimates of
  Climate Sensitivity, Earth's Future, 7, 677--690,
  \doi{https://doi.org/10.1029/2018EF001082}, 2019.

\bibitem[{Vega-Westhoff et~al.(2020)Vega-Westhoff, Sriver, Hartin, Wong, and
  Keller}]{Vega-Westhoff_paper}
Vega-Westhoff, B., Sriver, R.~L., Hartin, C., Wong, T.~E., and Keller, K.: The
  Role of Climate Sensitivity in Upper-Tail Sea Level Rise Projections,
  Geophysical Research Letters, 47, e2019GL085\,792,
  \doi{https://doi.org/10.1029/2019GL085792}, e2019GL085792 2019GL085792, 2020.

\bibitem[{Walker et~al.(2013)Walker, Haasnoot, and
  Kwakkel}]{dynamic_adaptive_pathways_2}
Walker, W.~E., Haasnoot, M., and Kwakkel, J.~H.: Adapt or Perish: A Review of
  Planning Approaches for Adaptation under Deep Uncertainty, Sustainability, 5,
  955--979, \doi{10.3390/su5030955}, 2013.

\bibitem[{Wang et~al.(2015)Wang, Lai, Chen, Yang, Zhao, and
  Bai}]{similar_work_3}
Wang, Z., Lai, C., Chen, X., Yang, B., Zhao, S., and Bai, X.: Flood hazard risk
  assessment model based on random forest, Journal of Hydrology, 527,
  1130--1141, \doi{https://doi.org/10.1016/j.jhydrol.2015.06.008}, 2015.

\bibitem[{Wong et~al.(2017{\natexlab{a}})Wong, Bakker, and
  Keller}]{AIS_disintegration_3}
Wong, T.~E., Bakker, A. M.~R., and Keller, K.: Impacts of Antarctic fast
  dynamics on sea-level projections and coastal flood defense, Climatic Change,
  144, 347--364, \doi{10.1007/s10584-017-2039-4}, 2017{\natexlab{a}}.

\bibitem[{Wong et~al.(2017{\natexlab{b}})Wong, Bakker, Ruckert, Applegate,
  Slangen, and Keller}]{BRICK}
Wong, T.~E., Bakker, A. M.~R., Ruckert, K., Applegate, P., Slangen, A. B.~A.,
  and Keller, K.: BRICK v0.2, a simple, accessible, and transparent model
  framework for climate and regional sea-level projections, Geoscientific Model
  Development, 10, 2741--2760, \doi{10.5194/gmd-10-2741-2017},
  2017{\natexlab{b}}.

\end{thebibliography}

    \end{nolinenumbers}%

\end{document}